\documentclass[reprint, superscriptaddress, secnumarabic, amssymb, nobibnotes, aps, prl]{revtex4-1}

\usepackage{graphicx}
\usepackage{epstopdf}
\usepackage[T1]{fontenc}
\usepackage[latin9]{inputenc}
\usepackage{amsbsy}
\usepackage{gensymb}
\usepackage[T1]{fontenc}
\usepackage[latin9]{inputenc}
\usepackage{amsmath}
\usepackage{amssymb}
\usepackage{bbm}
\usepackage{braket}
\usepackage{xcolor}
\allowdisplaybreaks
\usepackage{graphicx}
\usepackage[colorlinks=true]{hyperref}  
\hypersetup{
    bookmarks=true,         
    unicode=false,          
    pdftoolbar=true,        
    pdfmenubar=true,        
    pdffitwindow=false,     
    pdfstartview={FitH},    
    pdftitle={Muon spin spectroscopy study of the noncentrosymmetric superconductor TaOs},    
    pdfauthor={D. Singh, Sajilesh K.P, A. D. Hillier, R. P. Singh},     
    pdfsubject={},   
    pdfcreator={},   
    pdfproducer={}, 
    pdfkeywords={} {} {}, 
    pdfnewwindow=true,      
    colorlinks=true,       
    linkcolor=blue, 
    citecolor=blue,        
    filecolor=magenta,      
    urlcolor=blue           
} 
\usepackage[normalem]{ulem}

\newcommand{\equref}[1]{Eq.~(\ref{#1})}

\newcommand{\figref}[1]{Fig.~\ref{#1}}

\renewcommand{\approx}{\simeq}

\begin{document}
\title{\textrm{Nodeless s-wave superconductivity in the $\alpha$-$Mn$ structure type noncentrosymmetric superconductor TaOs: A $\mu$SR study}}
\author{D. Singh}
\affiliation{Indian Institute of Science Education and Research Bhopal, Bhopal, 462066, India}
\author{Sajilesh K.P}
\affiliation{Indian Institute of Science Education and Research Bhopal, Bhopal, 462066, India}
\author{Sourav Marik}
\affiliation{Indian Institute of Science Education and Research Bhopal, Bhopal, 462066, India}
\author{P.K.Biswas}
\affiliation{ISIS facility, STFC Rutherford Appleton Laboratory, Harwell Science and Innovation Campus, Oxfordshire, OX11 0QX, UK}
\author{A. D. Hillier}
\affiliation{ISIS facility, STFC Rutherford Appleton Laboratory, Harwell Science and Innovation Campus, Oxfordshire, OX11 0QX, UK}
\author{R. P. Singh}
\email[]{rpsingh@iiserb.ac.in}
\affiliation{Indian Institute of Science Education and Research Bhopal, Bhopal, 462066, India}

\date{\today}
\begin{abstract}
\begin{flushleft}

\end{flushleft}
Noncentrosymmetric superconductors can lead to a variety of exotic properties in the superconducting state such as line nodes, multigap behavior, and time-reversal symmetry breaking. In this paper, we report the properties of the new noncentrosymmetric superconductor TaOs, using muon spin relaxation and rotation measurements. It is shown using the zero-field muon experiment that TaOs preserve the time-reversal symmetry in the superconducting state. From the transverse field muon measurements, we extract the temperature dependence of $\lambda(T)$ which is proportional to the superfluid density. This data can be fit with a fully gapped s-wave model for $\alpha$ = $\Delta(0)/k_{B}T_{c}$ = 2.01 $\pm$ 0.02. Furthermore, the value of magnetic penetration depth is found to be 5919 $\pm$ 45 \text{\AA}, which is consistent with the value obtained from the bulk measurements.
\end{abstract}
\maketitle

\section{Introduction}

Superconductors with lack of spatial inversion symmetry in crystal lattice have attracted much attention in condensed matter physics from both experimental and theoretical perspective due to their fascinating and unusual electronic states \cite{EBA}. In noncentrosymmetric (NCS) superconductors, the relaxed space-symmetry leads to internal electric-field gradients and hence to antisymmetric spin-orbit coupling (ASOC) \cite{LP,EIR}, which lifts the two-fold spin degeneracy of the conduction-band electrons, leading to mixed singlet and triplet components. This parity mixing in NCS superconductors instigates a wide variety of unconventional properties. Illustrative examples include spin-triplet pairing in Li$_{2}$(Pd,Pt)$_{3}$Si \cite{LP1,LP2,LP3}, upper critical field close to or exceeding the Pauli limiting field in systems such as: CePt$_{3}$Si \cite{EB}, K$_{2}$Cr$_{3}$As$_{3}$ \cite{ABK1}, Nb$_{0.18}$Re$_{0.82}$ \cite{ABK2}, (Ta, Nb)Rh$_{2}$B$_{2}$ \cite{EMC}, and multiple superconducting gaps in LaNiC$_{2}$ \cite{ADH} and (La,Y)$_{2}$C$_{3}$ \cite{SKY}.\\
Another unique characteristic of the specific complex order parameters in noncentrosymmetric superconductors is time-reversal symmetry (TRS) breaking. The presence of TRS breaking in a superconductor is very rare and has been observed only in a very limited number of superconductors e.g. Sr$_{2}$RuO$_{4}$ \cite{SRO1,SRO2}, UPt$_{3}$ \cite{UP1,UP2,UP3}, (U,Th)Be$_{13}$ \cite{UB}, (Pr,La)(Os,Ru)$_{4}$Sb$_{12}$ \cite{PS1,PS2,PS3}, PrPt$_{4}$Ge$_{12}$ \cite{PPG}, LaNiGa$_{2}$ \cite{LNG}, Y$_{5}$Rh$_{6}$Sn$_{18}$ \cite{YRS}, and Lu$_{5}$Rh$_{6}$Sn$_{18}$ \cite{LRS}. Noncentrosymmetric superconductors have been recognized as good candidates to search for broken TRS. To date, TRS breaking have been found in NCS superconductors, such as LaNiC$_{2}$ \cite{ADH}, Re$_{6}$X (X = Ti,Zr,Hf) \cite{DKP,RPS,DSJ}, locally noncentrosymmetric SrPtAs \cite{PKB} and La$_{7}$Ir$_{3}$ \cite{JAT}. This number is very small, if compared to the extent of NCS superconductors studied so far \cite{VKA1,VKA2,MSAD,EBC,PK,DAMA,JATB,AAA,MNW}.\\ 
The intermetallic Re$_{6}$X-NCS superconductors are of particular interest since most of the members in this family have been found to exhibit TRS breaking in the superconducting state \cite{DKP,RPS,DSJ}. Interestingly, when studied using the muon spin rotation/relaxation ($\mu$SR) technique, the TRS-breaking effects seems very similar in all the Re$_{6}$X binary alloys. The persistence and independency of the particular transition metal X, points to a key role-played by Re. In fact, the recent appearance of TRS breaking in pure centrosymmetric Re indeed suggest that a lack of inversion symmetry is inessential and the local electronic structure of Re is crucial for the understanding of TRS breaking in Re$_{6}$X \cite{TMC}. One particularly intriguing proposal is that the broken TRS arises in Re$_{6}$X compounds due to the formation of Loop-Josephson-Current (LJC) state built on site, intraorbital, singlet pairing \cite{SG}. However, these recent results pose the more important question, which has not yet been resolved, namely why such energetics that would drive such state occur only in systems with Re and not in other elements.  However, to truly probe the origin of TRS breaking in Re$_{6}$X, a systematic study of additional noncentrosymmetric superconductors families, particularly Re-free $\alpha$-$Mn$ structure materials, is of great importance.\\
To this end, we study the binary transition metal compound TaOs, which exists in the same space group as Re$_{6}$X. Superconducting transition in TaOs appears around T$_{c}$ $\approx$ 2.1 K \cite{TO}. $\mu$SR study of Re-free $\alpha$-$Mn$ structure TaOS is essential as it provides an excellent opportunity to identify the origin of TRS breaking in the Re$_{6}$X NCS superconductors. Therefore, the superconducting state of TaOs was examined using the combination of zero-field muon measurements (ZF-$\mu$SR) and transverse-field muon measurements (TF-$\mu$SR). From ZF-$\mu$SR, we can infer about the state of TRS whereas TF-$\mu$SR allows estimating the symmetry of superconducting order parameter.

\begin{figure}
\includegraphics[width=1.0\columnwidth]{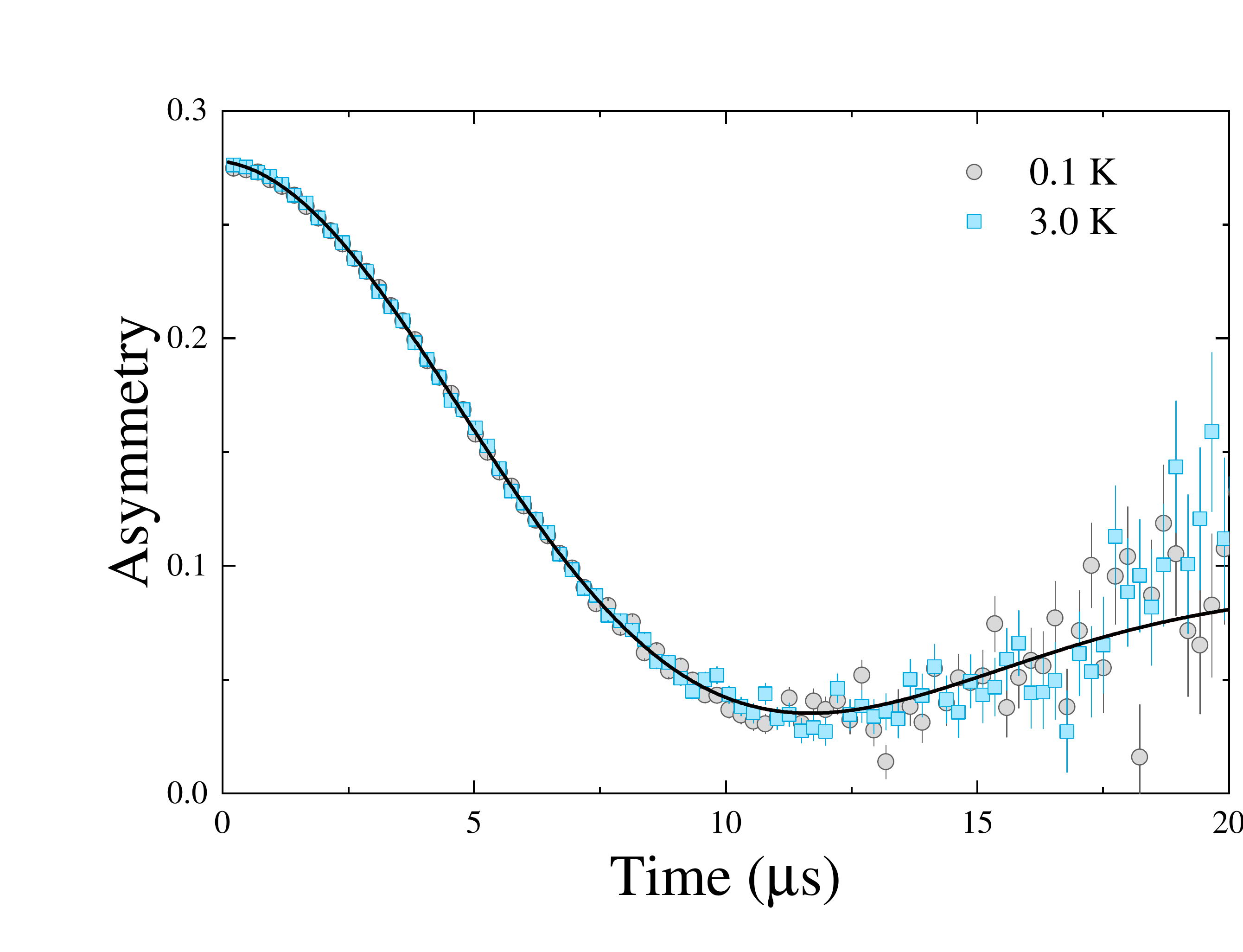}
\caption{\label{Fig1:ZFM}Zero-field $\mu$SR spectra collected below (0.1 K) and above (3 K) the superconducting transition temperature. The solid lines are the fits to Gaussian Kubo-Toyabe (KT) function given in Eq. \eqref{Fig1:ZFM}.}
\end{figure}

\section{Experimental Details}
The preparation of the polycrystalline TaOs sample used in this work is described in Ref.\cite{TO}. Powder x-ray diffraction (XRD) data confirm that the sample has the $\alpha$-$Mn$ crystal structure (space group I-43m No. 217), with no impurity phases detected. Magnetization, specific heat, and $\mu$SR measurements indicate that TaOs is a bulk superconductor with T$_{c}$  = 2.1 $\pm$ 0.1 K.
The $\mu$SR measurements were performed using the MuSR spectrometer at the ISIS  pulsed muon facility, STFC Rutherford Appleton Laboratory, Didcot, United Kingdom \cite{PJC}. In the transverse field mode, an external magnetic field was applied perpendicular to the muon-spin direction. The magnetic field was applied above the superconducting transition temperature of the sample and then cooled it to the base temperature. Muon spin rotates with the applied magnetic field and depolarizes as a consequence of magnetic field distribution inside the sample. Data were also collected in zero-field mode, where the muon spin relaxation is measured with respect to time. In the zero-field geometry, the stray fields at the sample position due to neighboring instruments and the Earth's magnetic field are cancelled to within $\sim$ 1.0 $\mu$T using three sets of orthogonal coils and an active compensation system. A full description of the $\mu$SR technique may be found in \cite{MSM}. The powdered TaOs sample was mounted on a silver holder and placed in a sorption cryostat, which we operated in the temperature range 0.1 K - 4 K.

\begin{figure}
\includegraphics[width=1.0\columnwidth]{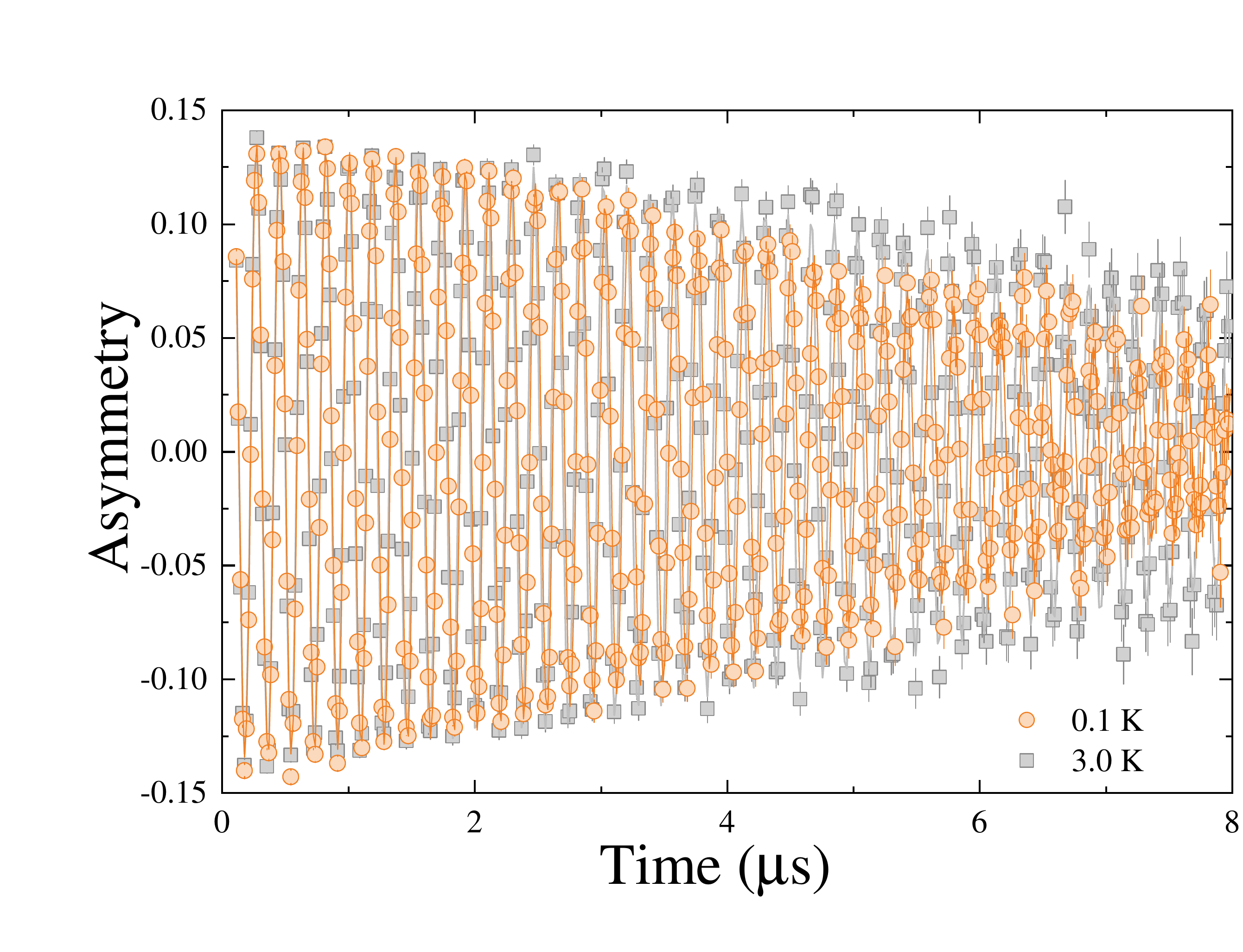}
\caption{\label{Fig2:TF} Representative TF-$\mu$SR signals collected at 3.0 K and 0.1 K in an applied magnetic field of 40 mT. The solid lines are fits using Eq. \eqref{eqn2:Tranf}.}
\end{figure}

\begin{figure*}[t]
\includegraphics[width=2.0\columnwidth]{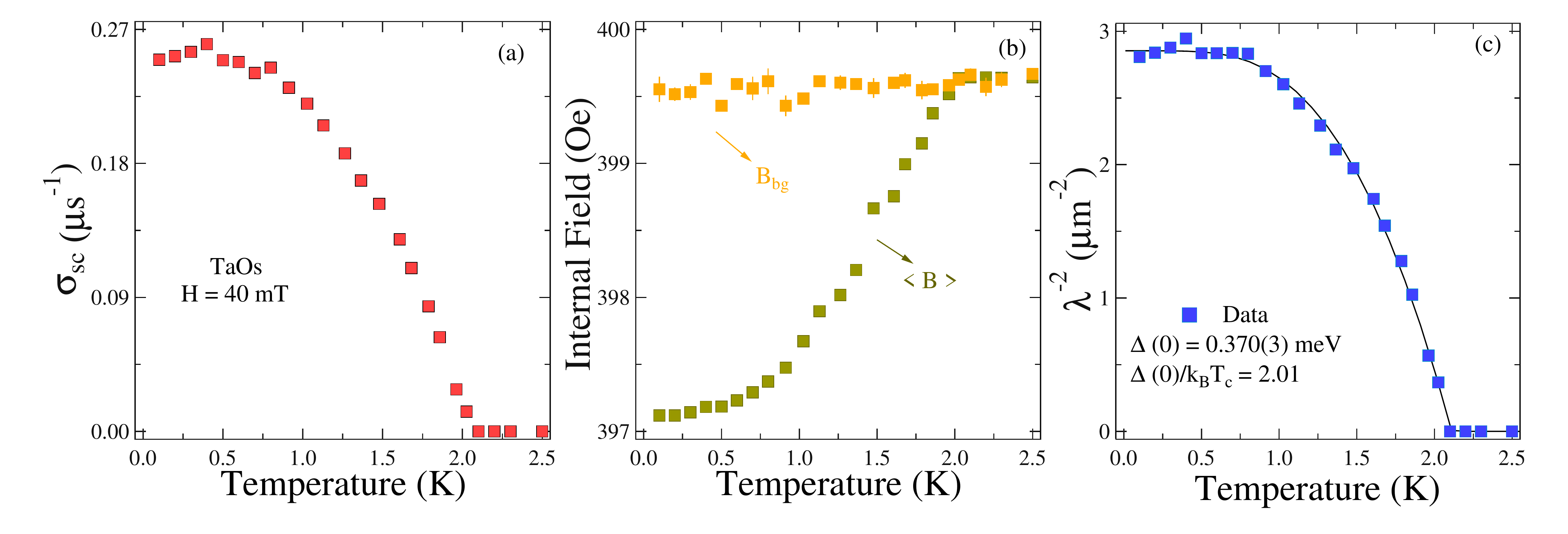}
\caption{\label{Fig3:sigma}(a) Temperature dependence of $\sigma_{\mathrm{sc}}$ collected at an applied field of 40 mT. (b) Temperature dependence of the internal magnetic fields experienced by the muon ensemble. <B> is the average magnetic field within the sample, whereas B$_{bg}$
is the field in the silver sample holder. (c) Temperature dependence of inverse square of the London penetration depth $\lambda^{-2}$. The solid line is the $s$-wave fit to the data.}
\end{figure*}

\section{Results and Discussion}

\subsection{Zero-field muon spin relaxation}
Firstly, time-reversal symmetry breaking was investigated by ZF-$\mu$SR, as has been seen in other NCS superconductors \cite{ADH,DKP,RPS,DSJ,PKB,JAT}. In superconductors with broken TRS, either the spin or orbital parts of the Cooper pairs are non-zero, which results in the appearance of the small magnetic field below the transition temperature. $\mu$SR can measure fields as small as 0.1 G, thus can measure the effect of TRS breaking in superconductors. Figure \ref{Fig1:ZFM} shows the ZF-$\mu$SR spectra measured at T = 0.1 K and 3.0 K, well below and above $T_{c}$. The absence of precessional signals suggests the absence of coherent internal fields which is generally associated with long-range magnetic ordering. There was no apparent change in the relaxation spectra in the superconducting state of TaOs, which suggests the absence of spontaneous internal magnetic fields. This confirms that time-reversal symmetry is preserved in the superconducting state of TaOs.\\
The time evolution of ZF-$\mu$SR spectra in the absence of atomic moments is best described by the Gaussian Kubo-Toyabe (KT) function \cite{RS} 
\begin{eqnarray}
G_{\mathrm{KT}}(t) &=& A_{1}\left[\frac{1}{3}+\frac{2}{3}(1-\sigma^{2}_{\mathrm{ZF}}t^{2})\mathrm{exp}\left(\frac{-\sigma^{2}_{\mathrm{ZF}}t^{2}}{2}\right)\right]\mathrm{exp}(-\Lambda t)\nonumber\\&+&A_{\mathrm{BG}} ,
\label{eqn1:zf}
\end{eqnarray} 
where $A_{1}$ is the initial asymmetry, $A_{\mathrm{BG}}$ is the time-independent background contribution from the muons stopped in the sample holder, $\sigma_{\mathrm{ZF}}$ is the muon spin relaxation rate, referring to the nuclear dipole moments and $\Lambda$ is the electronic relaxation rate.
For temperatures below and above $T_{c}$ the relaxation rate $\sigma_{\mathrm{ZF}}$ and $\Lambda$ does not change within the resolution of the instrument ($\Delta\Lambda$ $<$ 0.003 $\mu$s$^{-1}$ , $\Delta \sigma_{\mathrm{ZF}}$ $<$ 0.002 $\mu$s$^{-1}$). This indicates the absence of spontaneous magnetic fields in the superconducting state and suggests that the time-reversal symmetry is preserved, as would be expected in a conventional s-wave or a d-wave superconductor.

 \subsection{Transverse-field muon spin rotation} 
The TF-$\mu$SR data were collected after cooling the sample in an applied field of 40 mT (> $H_{c1}(0)$) from the normal state to the superconducting state. A set of typical measurements (below and above $T_{c}$) is shown in \figref{Fig2:TF}. Below $T_{c}$, the relaxation rate increases due to the presence of a non-uniform local field distribution as a result of the formation of a flux-line lattice in the superconducting state. The TF-$\mu$SR asymmetry spectra described by the sum of cosines, each damped with a Gaussian relaxation term: \cite{AMR,MWA}:
\begin{equation}
G_\mathrm{TF}(t) = \sum_{i=1}^N A_{i}\exp\left(-\frac{1}{2}\sigma_i^2t^2\right)\cos(\gamma_\mu B_it+\phi),
\label{eqn2:Tranf}
\end{equation}
where $A_{i}$ initial asymmetry, $\sigma_i$ is the Gaussian relaxation rate, $\gamma_{\mu}/2\pi$ = 135.5 MHz/T is the muon gyromagnetic ratio, common phase offset $\phi$, and $B_i$ is the first moment for the $i$th component of the field distribution. We found that the asymmetry spectra is described by two oscillating functions (N=2). In these fits, the i = 1 depolarization component was fixed to $\sigma_{1}$ = 0, which corresponds to a background term arising from those muons stopping in the silver sample holder as they do not appreciably depolarize over the time-scale of the experiment.\\
 The temperature dependence of the Gaussian depolarization rate was collected for several temperatures above and below $T_{c}$.  Figure \ref{Fig3:sigma}(a) shows the deploarization contribution from the superconducting state, which is calculated from the total $\sigma$ by subtracting the background contribution as $\sigma_{\mathrm{sc}} = \sqrt{\sigma^{2} - \sigma_{\mathrm{N}}^{2}}$. Below $T_{c}$ $\approx$ 2.1 K, the $\sigma_{\mathrm{sc}}$ increases systematically as the temperature is lowered. The temperature dependence of the internal magnetic field at the muon site <B> and the background field in the silver sample holder B$_{bg}$ [see \ref{Fig3:sigma}(b)] has also been extracted. The expected diamagnetic shift below $T_{c}$ is evident as a reduction of average internal field, while the B$_{bg}$ is approximately constant over the entire temperature range. Above $T_{c}$, the measured field corresponds to the applied field, which has an average value of (399.5 $\pm$ 0.5) Oe.

\begin{figure}
\includegraphics[width=1.0\columnwidth]{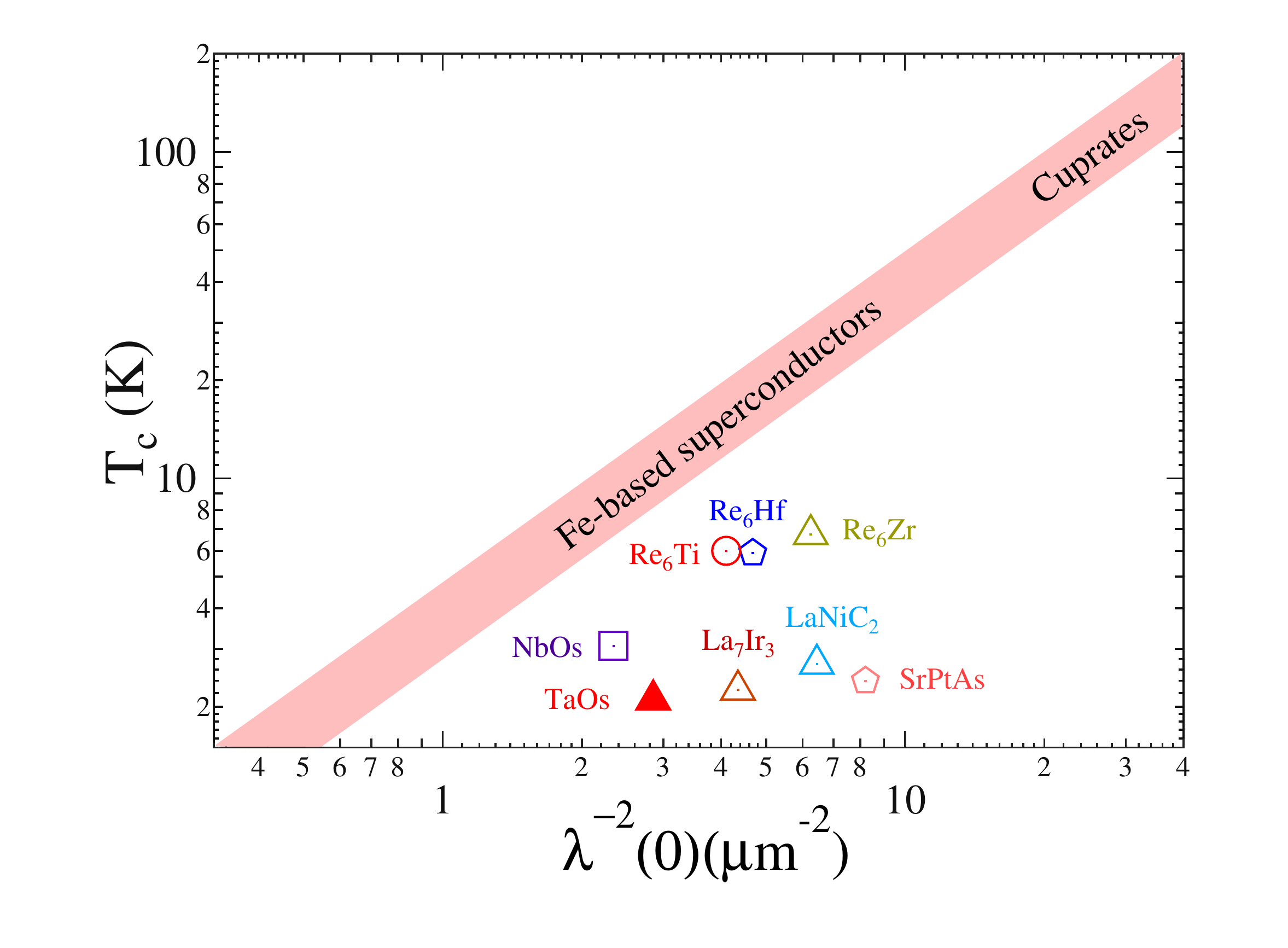}
\caption{\label{Fig4:UP} The Uemura plot showing the superconducting transition temperature $T_{c}$ vs the inverse squared penetration depth $\lambda^{-2}$(0), where TaOs is shown as a red solid triangle outside the range of band of unconventional superconductors. Other data points are the different NCS superconductors with broken TRS \cite{ADH,DKP,RPS,DSJ,JAT,PKB}.}
\end{figure}

In an isotropic type-II superconductor, the magnetic penetration depth $\lambda$ is related to $\sigma_{\mathrm{sc}}$ by the equation \cite{EH}:
\begin{equation}
\sigma_{\mathrm{sc}}(\mu s^{-1}) = 4.854 \times 10^{4}(1-h)[1+1.21(1-\sqrt{h})^{3}]\lambda^{-2}, 
\label{eqn3:sigmaH}
\end{equation}
where $h = H/H_{c2}$ is the reduced field. The above equation is valid for systems $\kappa > 5$. For TaOs, $\kappa = 56$ and temperature dependence of $H_{c2}$ was used from Ref. \cite{TO}. The temperature dependence of $\lambda^{-2}$ was extracted by incorporating data of $\sigma_{\mathrm{sc}}$ and $H_{c2}$(T) in \equref{eqn3:sigmaH}, as displayed in  \figref{Fig3:sigma}(c). The solid line in \figref{Fig3:sigma}(c) represents the temperature dependence of the London magnetic penetration depth $\lambda(T)$ within the local London approximation for an isotropic gap BCS superconductor in the dirty limit \cite{MHF}:
\begin{equation}
\frac{\lambda^{-2}(T)}{\lambda^{-2}(0)} = \frac{\Delta(T)}{\Delta(0)}\mathrm{tanh}\left[\frac{\Delta(T)}{2k_{B}T}\right] ,
\label{eqn4:lpd}
\end{equation}
where $\Delta(T)/\Delta(0) = \tanh\{1.82(1.018({T_{c}/T}-1))^{0.51}\}$ \cite{AC,HP} is the BCS approximation for the temperature dependence of the energy gap. The dirty-limit expression used because of $l<<\xi$, which is consistent with the TaOs \cite{TO}. By fitting the above discussed model it yields energy gap $\Delta (0)$ = 0.37 $\pm$ 0.02  meV and $T_{c}$ = 2.11 $\pm$ 0.02  K . The gap to $T_{c}$ ratio $\Delta (0)/k_{B}T_{c}$ = 2.02 $\pm$ 0.12, which is much higher than the value of 1.764 expected from the BCS theory, suggesting that TaOs is a strongly-coupled superconductor. The penetration depth at T = 0 obtained from the present microscopic study $\lambda$ (0)  =   5919 $\pm$ 45 \text{\AA}, is consistent with the value derived from the bulk properties in TaOs, $\lambda$(0) = 5168 \text{\AA}\cite{TO}.\\
Figure \ref{Fig4:UP} shows the Uemura plot where the shaded region depicts the "band of unconventionality," consisting of many unconventional superconducting systems, such as Fe-based superconductors and high-temperature superconductors \cite{YJ1,YJU}. We included NCS superconductors with broken TRS in the plot, using the published data for the related compounds \cite{ADH,DKP,RPS,DSJ,JAT,PKB}. TaOs with $\alpha$-$Mn$ structure is positioned well outside the broad line for unconventional superconductors suggesting that the superconducting mechanism is primarily conventional. Interestingly, Re$_{6}$X compounds occupy the same region in the Uemura plot, probably indicating that the underlying mechanism of the TRS breaking in this family is likely to be similar. It should be noted that theoretically it is suggested that the extent of ASOC determines the pairing mixing ratio. However, it was found that this is not the case in Re$_{6}$X. For example, in Re$_{6}$X(X = Ti,Zr,Hf) family, replacing the 3$d$ atoms with the 5$d$ atoms at the X-sites, does not show any discernible difference in the TRS breaking signal, as confirmed by the comparable magnitude of the mean internal field $\Delta B_{int}$ \cite{DKP}. Shang et al., has shown that the $B_{int}$ in the $\alpha$-$Mn$ type Re-based compounds scales linearly with the nuclear magnetic moment $\mu_{N}$ \cite{TMC}. The appearance of TRS breaking in pure Re indicates that a lack of inversion symmetry is inessential. More importantly, taking into account the case of Re$_{6}$X (X = Ti,Zr,Hf) and invariance of TRS in OsX (X = Nb,Ta) (Re-free), strongly supports the conclusion of key role played by Re electronic structure in the origin of TRS breaking in the Re$_{6}$X family as previously predicted in Ref. \cite{TMC}.
\section{Summary}
We present the muon spin rotation/relaxation studies of the binary compound TaOs. ZF-$\mu$SR results show the absence of any spontaneous magnetic fields below $T_{c}$, suggesting that the time-reversal symmetry is preserved in this system. The absence of TRS breaking in TaOs supports the conclusion that the Re-electronic structure is playing a key role in the observed TRS breaking in Re$_{6}$X family. Also, the recent observation of TRS breaking in pure Re metal further validates the above conclusion. Notably, symmetry analysis in Ref.\cite{SG} also proposes that Re$_{6}$X satisfies the exotic instability of the LJC ordered superconducting state and can be the possible reason for TRS breaking in these family of compounds. However, it is surprising why such energetics if present occurs only for systems with Re and why not for system crystallizing in the same $\alpha$-$Mn$ structure. Meanwhile, the temperature dependence of $\lambda(T)$ was determined from the TF-$\mu$SR, which is well described by an isotropic gap BCS model with a $\Delta(0)/k_{B}T_{c}$ = 2.02 $\pm$ 0.12, revealing strongly-coupled superconductivity in TaOs. The magnetic penetration depth found to be $\lambda$ = 5919 $\pm$ 45 \text{\AA}, which is close to the value reported earlier by magnetization data. Furthermore, in Uemura plot, TaOs is placed well outside the broad line for unconventional superconductors, indicating a conventional mechanism of superconductivity. Altogether, the combined results of $\mu$SR studies suggest that TaOs is a strongly-coupled s-wave superconductor. 
\section{Acknowledgment}

R.~P.~S.\ acknowledges Science and Engineering Research Board, Government of India for the Young Scientist Grant No. YSS/2015/001799 and Department of Science and Technology, India (SR/NM/Z-07/2015) for the financial support and Jawaharlal Nehru Centre for Advanced Scientific Research (JNCASR) for managing the project. We thank ISIS, STFC, UK for the beamtime to conduct the $\mu$SR experiments \cite{doi}.

\end{document}